\documentclass[pdflatex,sn-nature]{sn-jnl}

\usepackage{hyperref}
\usepackage{graphicx}%
\usepackage{multirow}%
\usepackage{amsmath,amssymb,amsfonts}%
\usepackage{amsthm}%
\usepackage{mathrsfs}%
\usepackage[title]{appendix}%
\usepackage{xcolor}%
\usepackage{textcomp}%
\usepackage{manyfoot}%
\usepackage{booktabs}%
\usepackage{algorithm}%
\usepackage{algorithmicx}%
\usepackage{algpseudocode}%
\usepackage{listings}%
\usepackage{braket}
\usepackage{todonotes}
\usepackage{lineno}
\usepackage{url}
\usepackage{multicol}
\usepackage{tabularx}
\usepackage{geometry}
\theoremstyle{thmstyleone}%
\theoremstyle{thmstyletwo}%

\theoremstyle{thmstylethree}%
\newcommand{\figref}[1]{Fig.\ \ref{#1}}

\begin{document}

\title{Multiband Topological Heterojunctions on the Surface Nanoscale Axial Photonics Platform}


\author[1,2]{\fnm{Nathaniel} \sur{Fried}}\email{friednr@umd.edu}

\author*[2]{\fnm{Dashiell L. P.} \sur{Vitullo}}\email{dashiell.l.vitullo.civ@army.mil}

\author*[1,3,4]{\fnm{Avik} \sur{Dutt}}\email{avikdutt@umd.edu}

\affil[1]{\orgdiv{Institute for Physical Science and Technology}, \orgname{University of Maryland}, \orgaddress{\street{4254 Stadium Drive}, \city{College Park}, \postcode{20742}, \state{MD}, \country{USA}}}

\affil[2]{\orgdiv{DEVCOM} \orgname{Army Research Laboratory}, \orgaddress{\street{2800 Powder Mill Rd}, \city{Adelphi}, \postcode{20783}, \state{MD}, \country{USA}}}

\affil[3]{\orgdiv{National Quantum Laboratory (Qlab)}, \orgname{University of Maryland}, \orgaddress{\street{4505 Campus Drive}, \city{College Park}, \postcode{20740}, \state{MD}, \country{USA}}}

\affil[4]{\orgdiv{Department of Mechanical Engineering}, \orgname{University of Maryland}, \orgaddress{\street{4298 Campus Dr}, \city{College Park}, \postcode{20742}, \state{MD}, \country{USA}}}

\abstract{Analogue Hamiltonian simulation (AHS) in photonic systems can be an enticing alternative to direct experimental study of complex Hamiltonian systems as a result of the low cost and high degree of control one can have over the system's properties. Notably, the field of topological photonics has emerged in the last decade primarily by simulating tight-binding models of electrons within topologically nontrivial condensed-matter systems. Optical simulation of topologically nontrivial Hamiltonians requires optical resonators with minimal loss and well-matched frequencies whose intersite coupling can also be precisely controlled. The Surface Nanoscale Axial Photonics (SNAP) platform \cite{sumetsky_surface_2011, Sumetsky2013, sumetsky_optical_2019} satisfies all these requirements, exhibiting ultra-low loss operation and sub-angstrom fabrication precision, making it an excellent platform for AHS. In this work, we experimentally demonstrate the first topologically nontrivial photonic SNAP devices by coupling together axial modes of adjacent SNAP microresonators to form a variety of Su-Schrieffer-Heeger (SSH) lattices \cite{su_solitons_1979}. The devices manifest numerous distinct topological band structures corresponding to each axial mode of the microresonators, enabling us to observe behavior both close to and far from the topological-trivial phase transition. We further expand the scope of topological SNAP systems to contain not just higher-order generalizations of SSH lattices, but junctions between multiband lattices with dissimilar topological phases created by coupling up to 21 uniform and well-matched SNAP microresonators. Analyzing such ``heterojunctions" necessitated our development of generalized topological polarization methods. We thus demonstrate the exceptional promise of the SNAP platform for AHS of 1D topological insulators, and also open the door to the potential for simulating $>$2 dimensional systems by utilizing nonlinear interactions.}
\maketitle
\section{Introduction}\label{sec:intro}\vspace{-10pt}\begin{multicols}{2}
Analogue Hamiltonian simulation (AHS) has emerged as a powerful paradigm to study the nontrivial physics -- in particular topological physics -- hosted by a plethora of lattice models \cite{georgescu_quantum_2014, cubitt_universal_2018, lloyd_universal_1996, mauron_predicting_2025}. By engineering the couplings between the lattice sites of an easily controlled physical system (the simulator) to mimic the couplings in the Hamiltonian of another, more difficult system (the simulated system), AHS enables the study of the simulated system with a relaxed set of experimental restrictions \cite{daley_twenty-five_2023}. This approach has seen rapid advances across several quantum platforms such as superconducting circuits \cite{lamata_digital-analog_2018,houck_-chip_2012}, trapped ions \cite{davoudi_towards_2020,dumitrescu_dynamical_2022}, and photonic circuits \cite{tan_photonic_2014,luo_quantum_2015}, each of which offers distinct advantages in control, scalability, and implementation. Hamiltonians with topological properties are of particular interest, as those properties can be simultaneously robust to local fluctuations \cite{peterson_quantized_2018} but sensitive to global effects \cite{jin_yin_ni_soljacic_zhen_peng_peng_2019}. For example, topological systems have been used to demonstrate back-scattering immune transport \cite{wang_observation_2009,roushan_topological_2009} and to measure fundamental constants with unmatched precision \cite{jeckelmann_quantum_2001}. Several key factors determine a particular platform's suitability for simulating topologically nontrivial phenomena: Most obviously, it must be possible to manipulate the simulator system in such a way that the dynamics of its degrees of freedom align with those of the simulated system. But also, the simulator must accurately mimic the dynamics as opposed to succumbing to loss or decoherence as a result of nonideal factors like fabrication imperfections, noise, or unintentional interactions. 

Photonic devices are among the most common platforms used to perform AHS \cite{aspuru-guzik_photonic_2012} of topologically nontrivial lattice models \cite{price_roadmap_2022,tan_photonic_2014,luo_quantum_2015,yang_simulating_2022,saxena_photonic_2022} as a result of the photons' mobility and lack of interactions relative to other platforms. Both of these features enable photons to traverse large optical microresonator lattices or networks with minimal noise and loss \cite{kitagawa_observation_2012,broome_photonic_2013,broome_discrete_2010, saxena_photonic_2022}. The simulators take the form of either arrays of coupled microresonators \cite{sharp_near-visible_2024} or waveguide arrays \cite{rechtsman_photonic_2013,shen_floquet_2023} where the propagation direction takes the role of time. Additionally, naturally independent photonic degrees of freedom, e.g.  frequency or orbital angular momentum, can often be effectively manipulated as synthetic dimensions \cite{arguello-luengo_synthetic_2024,yang_simulating_2022} to replace or complement the physical dimensions of the simulated system. Synthetic dimensions can improve the scalability of optical systems \cite{senanian_programmable_2023}, and can also be used to analyze systems that are impossible to physically create since synthetic dimensions often lack restrictions like locality \cite{yuan_synthetic_2018}, unitarity \cite{leefmans_topological_2022}, or dimensionality \cite{dutt_higher-order_2020,liu_multidimensional_2023}. However, one primary challenge of photonic AHS devices is the requirement that microresonator arrays be both low in loss (and thus narrow in linewidth), and well-matched in frequency, which often necessitates active optical elements \cite{saxena_realizing_2022, flower_observation_2024}.

\begin{figure}[H]\label{fig:1}
    \centering
    \includegraphics[width=0.9\linewidth]{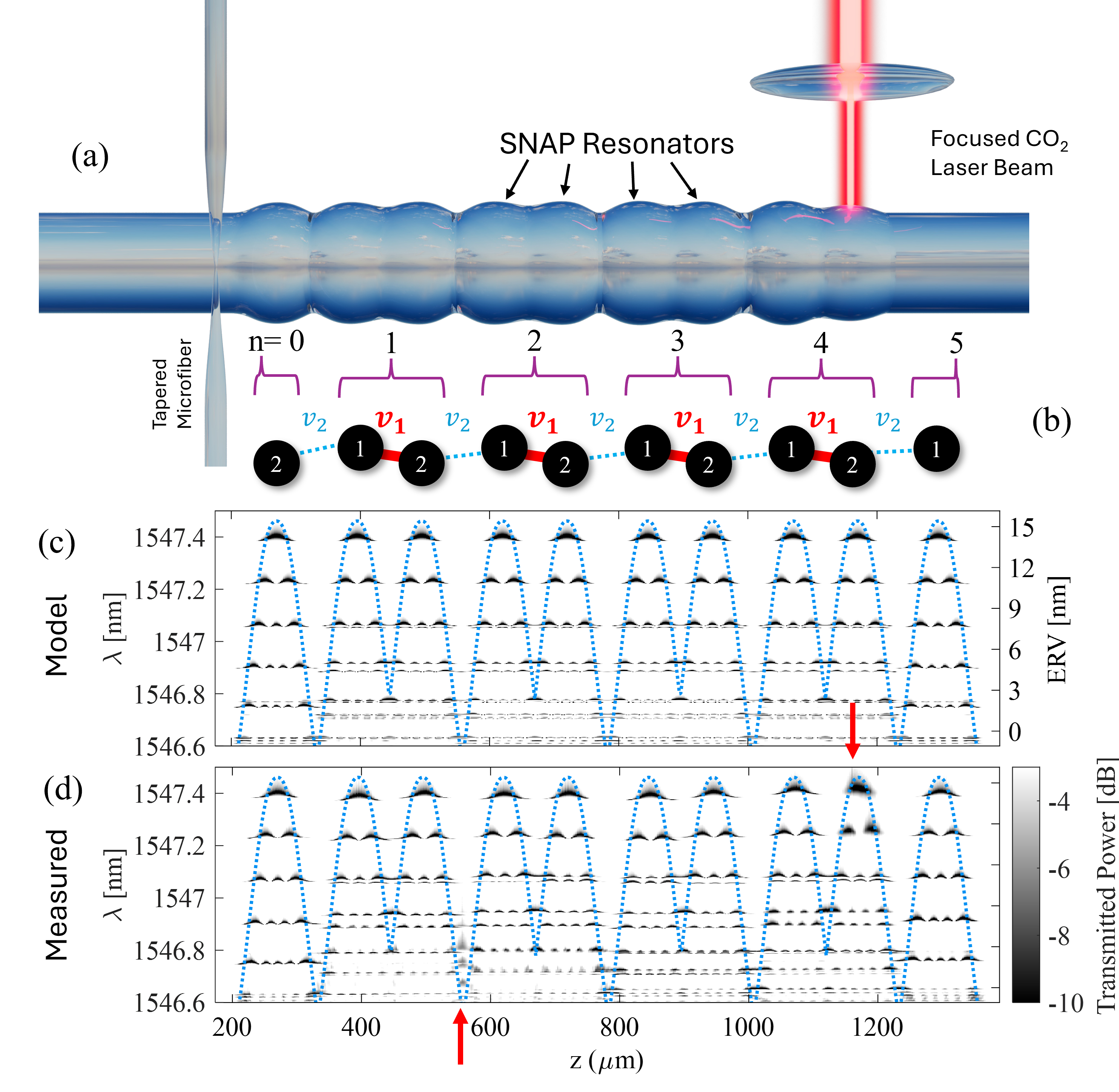}
    \caption{\textbf{(a)} Visualization of the CO$_2$ laser annealing fabrication process and the tapered optical microfiber scanning method (feature sizes not to scale). \textbf{(b)} A ``polymer'' depiction of an SSH2 lattice with 10 sites and staggered couplings $v_1$ and $v_2$. Unit cells are labeled by an integer $n\in \{0,1,\ldots,5\}$, with the depicted structure having 4 full unit cells and 2 half unit cells. Each of the sites within are labeled by an integer $s=1,2$. Spectrograms from a \textbf{(c)} computational model and \textbf{(d)} experimental data  of the above SNAP device presenting several axial orders corresponding to the band structure of lattices analogous to that of (b). Blue lines represent the estimated ERV profile along the fiber axis and thus an optical potential. Red arrows point to locations of loss-inducing localized defects.}
\end{figure}

The Surface Nanoscale Axial Photonics (SNAP) platform has emerged as a promising option to tackle this challenge. SNAP devices are made by engineering nanoscale effective radius variations (ERV) into the air-glass interface of the surface of optical fibers [\figref{fig:1}(a)]  \cite{sumetsky_optical_2019}. They demonstrate extremely low propagation-losses ($<0.001$ dB/cm) \cite{sumetsky_coupled_2012,sumetsky_surface_2011} and intrinsic quality factors beyond $10^8$ \cite{pollinger_all-optical_2010}. SNAP shows substantial advantages in equipment costs and fabrication speed compared to integrated photonic systems often used in investigations of topological photonics \cite{sharp_near-visible_2024}. For example, a single optical fiber can host dozens of identical SNAP resonators with sub-angstrom wavelength precision \cite{toropov_permanent_2016a} since SNAP microresonators can easily be post-processed after characterization. This post-processing or trimming is non-volatile and local to each microresonator, that is, it does not disturb other parts of the device \cite{sumetsky_snap:_2012}, and has negligible additional manufacturing costs compared to integrated photonic chips fabricated by planar lithography. Hence, it is natural to ask whether topological photonics can be experimentally investigated on the SNAP platform. 

Here we answer this question in the affirmative by reporting the first SNAP-based demonstrations of several topologically nontrivial Hamiltonians. Furthermore, we leverage the speed and relative ease with which new SNAP devices can be fabricated to showcase a more complex and less-studied structure -- a junction between distinct higher-order Su-Schrieffer-Heeger (SSH) lattices \cite{su_solitons_1979}. While this work is focused on 1D Hamiltonians, the small mode volume and low losses of SNAP resonators enable nonlinear effects to manifest at low intensities \cite{pollinger_all-optical_2010}. This opens the possibility for both the axial quantum numbers and the non-axial quantum numbers to be utilized as a synthetic dimension to implement two- or higher-dimensional topological models.\end{multicols}
\begin{multicols}{2}

\section{Implementation of SNAP SSH Lattices}
SSH lattices \cite{su_solitons_1979,xie_topological_2019,anastasiadis_bulk-edge_2022,verma_bulk-boundary_2024,lee_winding_2022,pineda_edge_2022,alexandradinata_wilson-loop_2014,sridhar_measuring_2025,rhim_bulk-boundary_2017,marques_one-dimensional_2019,wielian_transfer_2025,hotte-kilburn_integrated_2025, st-jean_lasing_2017, pellerin_wave-function_2024} consist of 1D arrays of sites $\ket{z_j}$ for $j=1, 2, 3, \ldots, L$ which are coupled in a local and periodic manner (we refer to a lattice with unit cells of length $\ell$ as an SSH$\ell$ lattice), and are hallmarked by their topologically protected edge modes \cite{zhao_interface_2024,mandal_topological_2024,pineda_edge_2022,shen_floquet_2023}. The general Hamiltonian of the SSH-like lattices we consider can be written as:
\vspace{-8pt}\begin{equation}
   \hat H = \sum_{j=1}^L u(j)\ket{z_j}\bra{z_j}+v(j) \Big(\ket{z_{j+1}}\bra{z_j}+\ket{z_j}\bra{z_{j+1}}\Big),
\end{equation}\label{eqn:Hamiltonian}
where $v(j)=v(j+\ell)$ are the adjacent couplings and $u(j)=u(j+\ell)$ are self-couplings. While SSH models only consider the ideal case that $u(j)=0$ or constant, we include them while analyzing our Hamiltonian models as they will appear unintentionally and require additional effort to compensate for them. While the conventional 2-band SSH has been extensively explored theoretically and experimentally \cite{dutt_higher-order_2020,sridhar_measuring_2025,hotte-kilburn_implementation_2024}, less work has focused on its $\ell$-band extensions with $\ell>2$ \cite{xie_topological_2019,anastasiadis_bulk-edge_2022,verma_bulk-boundary_2024}.
Fig.\ \ref{fig:1} depicts the relationship between (a) the fabrication of a SNAP device, and (b) the abstract 1D lattice model instantiated by the completed device. Annealing a section of the fiber by CO$_2$ laser exposure induces positive ERV (dilations) \cite{sumetsky_optical_2019, sumetsky_surface_2011, sumetsky_snap:_2012, sumetsky_coupled_2012}, localizing a discrete series of optical modes to sites along the fiber's axis (Fig.\ \ref{fig:1}(a)). Adjacent resonators couple evanescently, which also couples and hybridizes their whispering gallery modes (WGMs). The unit cells resemble small molecules that are connected into polymer-like structures, motivating the analogous naming convention used in Fig.\ \ref{fig:1}(b), which depicts an SSH2 lattice, composed of 4 whole unit cells (referred to as ``dimers") and 2 incomplete unit cells on the edges (called ``monomers"). 

The optical spectrum of the resulting SNAP device is then probed by placing a tapered optical microfiber oriented perpendicular to the sample on the surface such that it is evanescently coupled to WGMs in the sample, and measuring the transmission spectrum through the tapered microfiber. The WGMs localized in the SNAP devices are modeled using  the Green's function of the SNAP system in the energy-position basis, with wavelength playing the role of energy (Eq.\ \eqref{eqn:1Dschrodinger}) \cite{sumetsky_theory_2012}. The bare Green's function models the WGM structure in the absence of any coupled tapers. The renormalized Green's function includes the effects induced by the presence of a taper upon transmission. Destructive interference causes transmission dips to occur when the input light is resonant with WGMs in the SNAP device. By resolving the transmission in both wavelength $\lambda$ and position $z$, the mode spectrum is obtained and visualized in spectrograms, with representative examples shown for the model and the experimental measurements in Fig.\ \ref{fig:1}(c) and \ref{fig:1}(d) respectively.

SNAP resonators support multiple axial modes simultaneously, each with a different evanescent tail extent beyond the microresonator, which allows the elegant realization of several SSH lattices with distinct coupling coefficients within a single 1D SNAP array. These axial modes, labeled by $q\in \{1,2,...\}$, exhibit Fano lineshapes \cite{Liminov:17}, corresponding to the multiple vertically offset transmission dips seen within each microresonator in the 1D array of Fig. \ref{fig:1}(c,d). The presence of the tapered microfiber renormalizes modal wavelengths and linewidths in proportion to the local modal intensity. Hence, the measured resonances most closely approximate the bare resonances (i.e.\ WGM resonances in the absence of any coupling to a tapered microfiber) near the zeros/nodes of the measured mode profiles. The ERV curve is estimated by fitting data from measured spectrograms to that of the computational model of the SNAP device \cite{Vitullo:20} as described in Appendix \ref{sec:fit}. 

The fitted ERV curve exhibits a highly uniform array, evidencing the precision with which we can fabricate a large number of SNAP devices through iterative characterization and localized post-processing (Sec.\ \ref{sec:exp}). On top of this uniform background, we observe localized distortions in the measured spectrogram near the positions $z=550$ $\mu$m and $z=1150 $ $\mu$m indicated by red arrows in Fig.\ \ref{fig:1}(d). Resonances that intersect those positions (particularly the $q=1, 2,$ and $4$ modes of the $n=4, s=2$ resonator) exhibit linewidth broadening. These are indicative of localized defects (e.g.\ an impurity such as dust on the surface of the SNAP device) inducing scattering losses; however, the defects did not cause notable changes to the edge modes, which attests to the topological protection they inherit from the underlying SSH lattice. 

To simulate a Hamiltonian with a system of coupled resonators, we analyze the behavior of its supermodes in relation to the modes of the individual resonators. By identifying the uncoupled modes of each resonator with the basis vectors of the lattice $\ket{z_j}$, perturbations to the optical system correspond to terms within a Hamiltonian, and the light's dynamics mimic time evolution under that Hamiltonian. 

Within a lone SNAP microresonator, for each polarization, azimuthal, and radial mode of the unperturbed fiber, there will be similar but independent axial mode series. The axial mode series can be determined by solving a 1D Schr\"odinger equation in the fiber's axial dimension $z$, where the ERV profile $\Delta r(z)$ plays the role of the potential, and the shift $\Delta \lambda_q$ of the $q^{\text{th}}$ axial mode from the cutoff wavelength $\bar \lambda$ plays the role of energy \cite{sumetsky_theory_2012}:
\begin{equation}
	\Bigg(-\frac{1}{2\beta_{0}^2}\frac{\partial^2}{\partial z^2}-\frac{\Delta r(z)}{\bar{r}}\Bigg)\psi^{q}(z)= -\frac{\Delta \lambda_q}{\bar\lambda} \psi^{q}(z). \label{eqn:1Dschrodinger}
\end{equation}
Here, $\beta_0$ is the longitudinal component of the material wavenumber, the base fiber radius is $\bar r$ at zero ERV, and $\psi^{q}(z)$ is the axial mode profile. In the tight-binding approximation \cite{chen_tight-binding_2021,rhim_bulk-boundary_2017}, the Hamiltonian of the SNAP microresonator array can be rewritten as:
\begin{equation} \label{eqn:expansion}
    \hat H(z', z)\approx\sum_{j',q'}\sum_{j, q}\ket{z_{j'}^{q'}} H_{q,j}^{q',j'}\bra{z_j^{q}},
\end{equation}
where the axial mode profiles and thus basis vectors $\ket{z^{q}_j}\equiv\psi^{q}(z-z_j)$ correspond to a fixed ERV profile $\Delta r(z-z_j)$, shifted to the location of the $j^{\mathrm{th}}$ resonator, and $H^{q',j'}_{q,j}$ are the matrix elements of the overall Hamiltonian in this basis. This methodology -- utilizing the bound modes of a complicated physical system to approximate idealized sites on a lattice -- can be considered a reversal of the widely used tight-binding approximation. In most tight-binding models, the Hamiltonian matrix elements can be determined directly by overlap integrals of the independent site potentials and their wavefunctions since the overall potential will simply be sum of the individual potentials \cite{chen_tight-binding_2021} corresponding to each ``atom" of the system. Since the laser annealing process features a nonlinear response to the laser dose \cite{sumetsky_snap:_2012}, the cumulative ERV profile after multiple nearby exposures will not follow the same rule (e.g.\ a single exposure typically creates a super-Gaussian ERV profile, but a subsequent nearby exposure with the same power and duration does not result in a system that is well modeled by the sum of two super-Gaussians). More details are in Appendix \ref{sec:fit} and in Figs.\ \ref{fig:B1}, \ref{fig:B2}.

\begin{figure}
    \centering
   \includegraphics[width=1\linewidth]{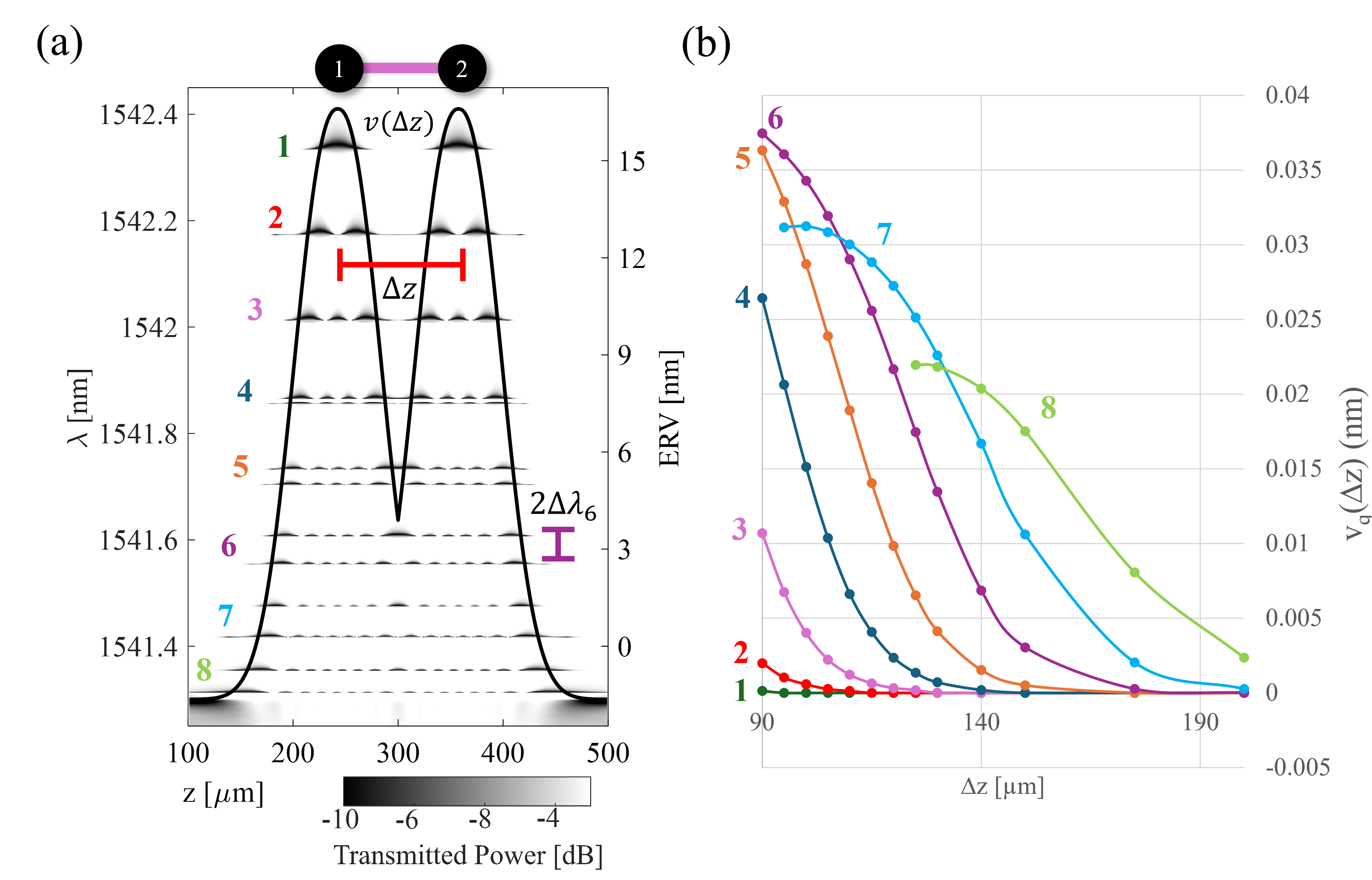}
    \caption{\textbf{(a)} Computed spectrogram of two adjacent sites with  $\Delta z =115$ $ \mu$m showcasing hybridization of axial modes (color coded). The $q=1$ fundamental modes are unhybridized, while the high-order modes are almost completely hybridized. \textbf{(b)} Wavelength splitting induced as a function of intersite spacing for $q=1$ to $8$ axial modes of the SNAP resonators. At large $\Delta z$, the expected exponential decay is observed. At small $\Delta z$, a saturation of the wavelength splitting, and hence of the coupling between the sites, is observed. }
    \label{fig:2}
\end{figure}

As a stepping stone on the path to building a full SSH lattice, we first investigate the effects of evanescent coupling between nearby resonators  by directly solving Eq.\ \eqref{eqn:1Dschrodinger} (see also Appendix \ref{sec:fit}). This 2-site computation allows us to estimate the tight-binding matrix elements between adjacent lattice sites as a function of the intersite spacing $\Delta z$. A characteristic spectrogram obtained by these computations is displayed in Fig.\ \ref{fig:2}. While the lower-order axial modes ($q\leq 2$, higher wavelength) remain effectively degenerate at the unperturbed wavelength $\lambda_q^{(0)}$, the higher order axial modes hybridize with their twin into symmetric and antisymmetric combinations with wavelengths of $\lambda_q^-$ and $\lambda_q^+$ respectively. Past the 5th axial order, the site potentials are mostly merged, and the only remaining feature signifying a separate ``left" and ``right" mode is the minute increases in the intensity near their midpoint. To first order, the coupling  between adjacent microresonators is given by the splitting of the resonances,  $  v^q (\Delta z)= (\lambda_q^+ - \lambda_{q}^-)/ 2$ and the self-couplings will be given by the shifting of their mean $u^q (\Delta z) = (\lambda_q^+ + \lambda_{q}^-)/{2}\, - \,\lambda_q^{(0)}$. This approximation is valid whenever the axial modes remain well separated in wavelength, i.e., when $\Delta z$ is much greater than the modal widths $\Delta w_q$. 
Fig.\ \ref{fig:2}(b) shows $v_q$ as a function of $\Delta z$ for each $q$, revealing how coupling strengths follow a decaying exponential when $\Delta z\gg\Delta w$, but saturate when $\Delta z\sim \Delta w_q$ (the curves are truncated after saturation as certain modeling assumptions explained in Appendix \ref{sec:fit} fail for $\Delta z < \Delta w_q$). $\Delta w_q$ increases with axial order, which explains why the higher order modes couple more strongly. Coupling strength increases for either high $q$ or small $\Delta z$, but very importantly, the \textit{variations} in coupling strength (i.e., their derivatives in respect to $\Delta z$) decrease in the same limit. We can thus expect the couplings between higher-order modes to be both stronger and more uniform than the lower order modes. For example, $v_5(100\,\mu m)/v_5(120\,\mu m)\sim 3$ but $v_6(100\,\mu m)/v_6(120\,\mu m)\sim 1.5$.
\end{multicols}
\begin{multicols}{2}
The couplings are nearly zero for spacings beyond $200$ $\mu$m, which allows us to ignore non-adjacent couplings throughout this work as the smallest adjacent spacings used are 100 $\mu$m. The only relevant terms within Eq.\ \eqref{eqn:expansion} are thus the self and adjacent couplings between modes of the same axial order ($q=q', |j-j'|\le 1$), i.e.\ those present in Eq.\ \eqref{eqn:Hamiltonian}. By controlling the spacing between sites in a lattice, the $v$ terms can be customized to any desired value between 0 and the saturation point, but not without inducing self-couplings. Therefore, a necessary consideration of making non-uniformly coupled resonator arrays is that the resonances will be detuned if we do not compensate for the associated nonuniform self-couplings. To compensate for the nonuniformity and also undo the effect of fabrication errors, we adjust the height of individual microresonators. This can be implemented with an increased initial dose from the CO$_2$ laser or by firing additional pulses during the post-processing step, as explained in Sec.\ \ref{sec:exp}. Note that if non-zero self-couplings were desired (e.g.\ to make lattices  with on-site detunings such as the Rice-Mele model \cite{chiel_fast_2024, asboth_short_2016}), one can use these steps to \textit{intentionally} induce self-couplings.

To summarize, when building a lattice from the $q^*$ axial mode of a SNAP microresonator array, the ${u}_{q^*}$ coefficient for each resonator can be chosen almost arbitrarily and the coupling coefficient ${v}_{q^*}$ between each pair of adjacent resonators can be chosen to be any positive value by using the coupling curves of Fig.\ \ref{fig:2}(b). For the axial mode lattices corresponding to $q$ greater than $q^*$, there will be less variations in both the ${v}_q$ and the ${u}_q$ coefficients. Therefore, the SNAP platform enables simultaneous investigation of multiple lattice configurations so long as the assumption that axial mode lattices are independent remains valid. 
\end{multicols}

\begin{figure}[htp]
    \centering
    \includegraphics[width=1\linewidth]{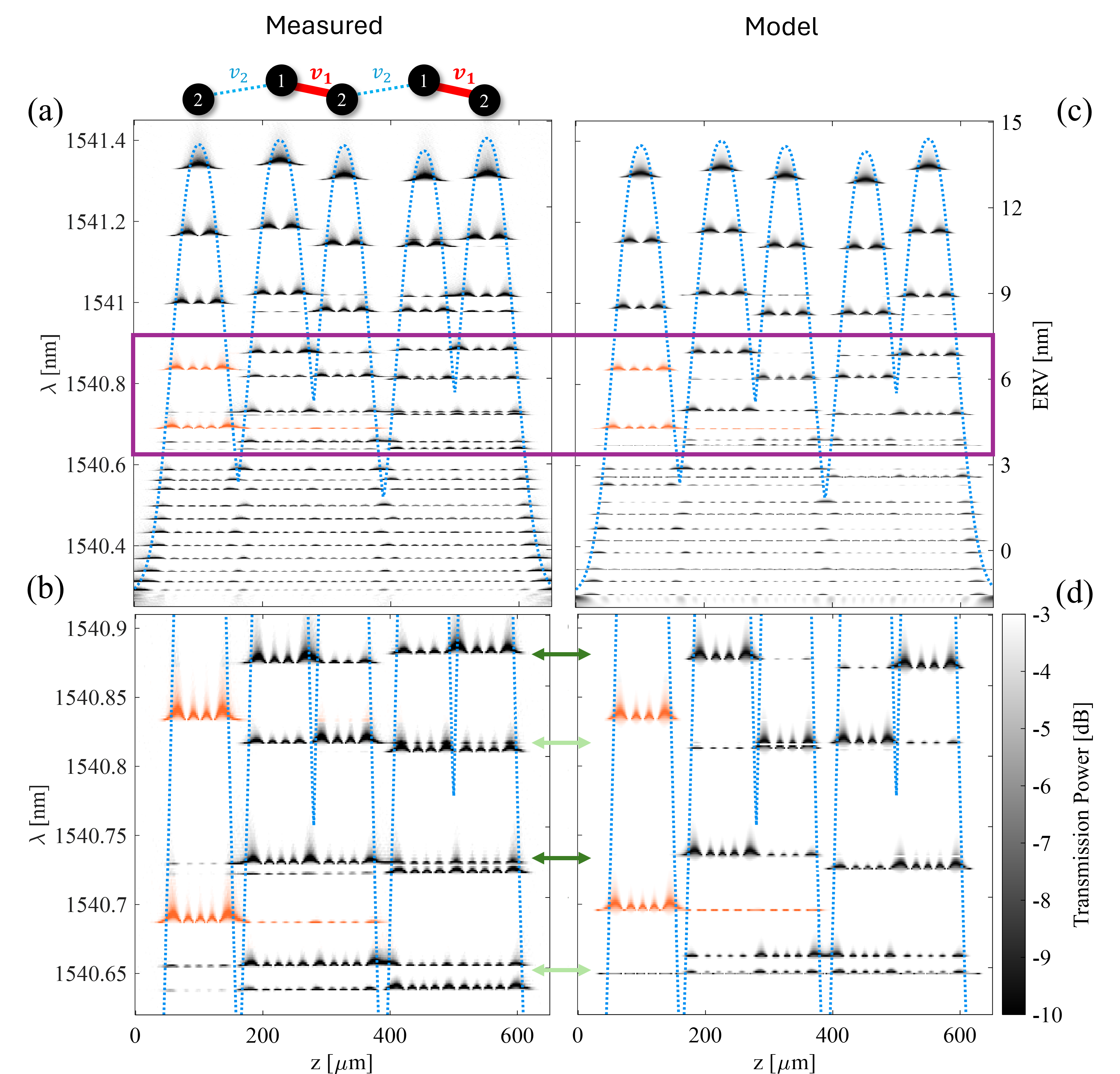}
    \caption{Measured (a,b) and model (c, d) spectrograms of 5-site SSH model. Purple rectangle corresponds to focused views (b,d) showing the $q=4$ and $q=5$ axial mode families. The lower-order mode families (higher in resonance wavelength) are too localized to observe coupling between the lattice sites. False color orange overlays highlight topological edge modes for $q=4,5$ and dark (light) green arrows indicate the positive (negative) energy bands. }     \label{fig:3}
\end{figure} 
\begin{multicols}{2}
\section{Experiment \& Results}\label{sec:exp}
In this work, the SNAP microresonators that serve as the building blocks of SSH lattices are initially formed using 30 ms pulses from a CO$_2$ laser operating at 5 W. The laser annealing of fused silica optical fibers induces momentary localized heating of the fiber surface through optical absorption \cite{sumetsky_surface_2011}. The fiber can thus recrystallize into a lower-stress state for which the effective radius (which depends on both the geometric radius and index of refraction) in the exposed section of the fiber is increased (Fig.\ \ref{fig:1}(a)). This thermally-induced stress release is inconsistent \cite{sumetsky_snap:_2012}, so it is difficult to predict the resulting ERV after any particular exposure. However, a unique benefit of the SNAP platform is the expedience with which SNAP devices can be characterized and post-processed to correct fabrication errors \cite{sumetsky_snap:_2012}. To ensure uniformity, the resonators are first scanned using a tapered microfiber to gauge the size and location of its imperfections, after which, the lowest ERV resonators are adjusted by firing additional laser pulses lasting between 20 and 25 ms. This process can be repeated until the microresonators are matched to within desired tolerance.

The measured spectrogram of a 5-site SSH2 (the standard SSH) structure is shown in Fig.\ \ref{fig:3}, accompanied by a best-fit computed spectrogram obtained using Eq.\ \ref{eqn:1Dschrodinger} with a potential as described in Appendix \ref{sec:fit}. It consists of 5 sites separated by alternating spacings of 125 $\mu$m and 100 $\mu$m, just like Fig. \ref{fig:1}. Notably, the parameters used for the resonators in Fig.\ \ref{fig:2} were obtained from fitting this data. This structure develops a single topological mode concentrated on the monomer whose intensity will decay exponentially as it leaks into the bulk at a rate of of $2\ln (v_1/v_2)$ per unit cell \cite{asboth_short_2016}.

We focus on the axial mode lattices with $q=4$ and $q=5$ (Fig.\ \ref{fig:3}(c,d)), where the key characteristics of the SSH model can be observed most directly.
As in Fig. \ref{fig:2}, the lattices of lower-order axial modes ($q \leq 3$) are effectively isolated, and for axial mode families with $q\geq 6$, the axial mode lattices can no longer be distinguished, as the potential wells have merged and the couplings have saturated.
In the $q=4,5$ axial mode lattices, the positive (red arrow) and negative (blue arrow) bands of modes are distributed throughout the bulk and separated by a bandgap. We observe edge modes that hallmark successful creation of topological models. While both topological edge modes are localized on the left side of the SNAP lattice, the $q=5$ edge mode was significantly less localized, such that its residual transmission after exponential decay from the edge can be seen on the right site of the $n=1$ cell around $z=300-400\ \mu$m. Based upon the locations of the resonances, we estimate that the couplings of the fifth modal order are approximately $v_2\approx 12$ pm and $v_1\approx 36$ pm, which are comparable to the predictions from Fig.\ \ref{fig:2} of $v_2=8$ pm and $v_1=30$ pm. The corresponding decay lengths are thus $(2\ln({v_1}/{v_2}))^{-1} \approx 0.46$ unit cells for the predicted couplings, and $\approx 0.38$ unit cells for the actual lattice. The small gap between the experimental couplings and the simulations' estimates can be attributed to a necessary assumption made in the fitting process that consistently underestimates the ERV profile between resonators (explained in Appendix \ref{sec:fit}). With this consideration, the SNAP device successfully replicates the physics expected from the SSH lattices.  
\end{multicols}
\begin{figure}[H]
    \centering
    \includegraphics[width=0.98\linewidth]{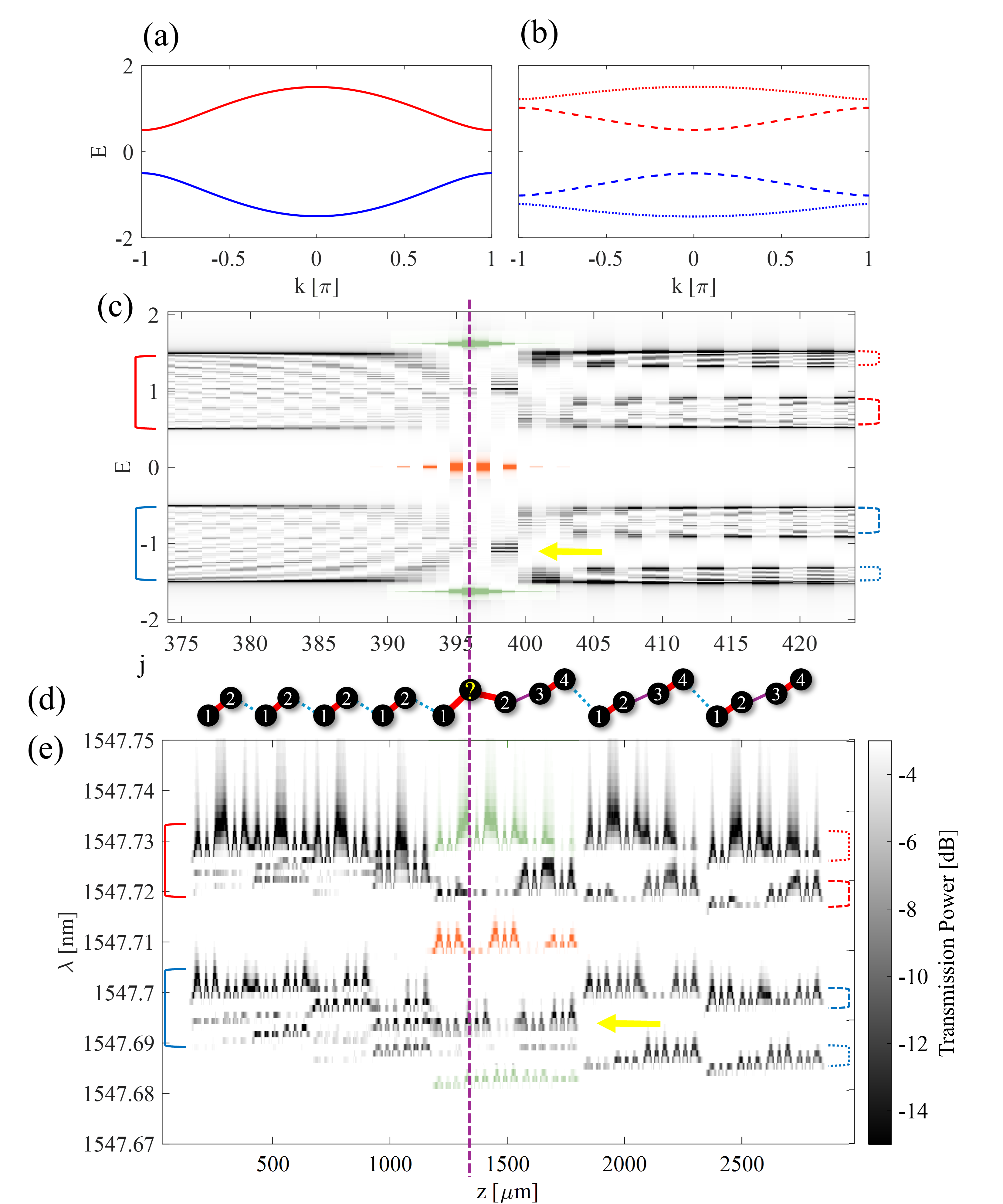}
    \caption{Band structure of a\textbf{ (a)} SSH2 lattice and of a \textbf{(b)} SSH4 lattice. Upper (lower) bands are red (blue), and inner (outer) bands are dashed (dotted). \textbf{(c)} bare tight-binding Green's function \cite{sumetsky_theory_2012} intensity (arb. units) of tight-binding model of a large SSH2 lattice on the left (${v}_{L}(s)=[1,0.5]$, $N=400$ unit cells) conjoined to an SSH4 lattice on the right (${v}_{R}(s)=[1,0.7,1,0.3]$, $N=400$ unit cells). Zero energy topological bound mode are tinted orange, non-zero energy bound modes are tinted green, and yellow arrows indicate the quasi-bound modes. Colored brackets next to the bands indicate the corresponding bands in (a,b). (d) Polymeric diagram of a 21 site lattice. The site on the (dashed purple) domain line can be considered a part of either lattice. (e) Measured spectrogram of a topological heterojunction. An  SSH2 lattice ($v_j$ correspond to spacings of 100 $\mu$m and $125$ $\mu$m) is conjoined to an SSH4 lattice ($v_j$ correspond to spacings of 100 $\mu$m, 105 $\mu$m, 100 $\mu$m, and 130 $\mu$m). This measured spectrogram closely follows the tight-binding spectrogram in (a) barring the expected local self-coupling shifts, also present in Figs.\ \ref{fig:1} and \ref{fig:3}. Arrows and brackets indicate correspondence between modes and bands as in (a-c).}
    \label{fig:4}
\end{figure}
\begin{multicols}{2}
Given the relative ease of producing the SNAP SSH lattices depicted in Fig.\ \ref{fig:1} and Fig.\ \ref{fig:3}, we next showcase the platform's versatility by instantiating the first heterojunction of higher-order (multiband) SSH lattices, which has not been accomplished on other platforms to the best of our knowledge. 
Specifically, we demonstrate an interface between an SSH2 section of a lattice and an SSH4 section. Fig.{} \ref{fig:4} shows the band structure of SSH4 lattices in comparison to SSH2 lattices (a,b), direct tight-binding simulations (c), and experiments (e), for which numerous localized modes emerged at the domain boundary. The overall lattice can be considered either as an 11 site SSH2 lattice attached to a 10 site SSH4 lattice, or a 10 site SSH2 lattice attached to an 11 site SSH4 lattice, since the allocation of the shared cite at the center is ambiguous. Either way, the last site of the SSH2 region is strongly coupled to the first unit cell of the SSH4 region, resulting in an effective ``pentamer" at the junction between the two regions. 

Fig.~\ref{fig:4}(c) shows the bare Green's function of a similar SSH lattice expressed in the site position $(j)$ basis rather than the real-space $(z_j)$ basis, computed by diagonalizing a large ($L=800$ sites) but finite tight-binding Hamiltonian. With no intrasite intensity variations and a wide bulk for the free modes to distribute themselves across, it is easy to spot the interface modes and compare them to the continuum modes. There are three real modes localized to this interface: a standard zero-energy mode in the central bandgap (orange), and  two nonzero-energy bound modes just outside the highest and lowest bands (green). While there are 4 bands on the right (SSH4) side, the auxiliary bandgaps are closed on the left (SSH2) side. Because of this, two quasi-modes (localized increases in spectral intensity spanning numerous modes as opposed to from a single mode) appear within the auxiliary bandgaps. We can view these quasi-modes as artifacts of the SSH4 lattice's auxiliary bandgap modes after having been coupled to the SSH2 lattice. While the topology of the right lattice forbids the auxiliary bandgap modes from entering it, the topology of the left side permits it, allowing leakage into its bulk. The small distance between the centers of the quasi-modes (localized on sites 3 and 4 of the pentamer) and the SSH2 side prevents them from being totally delocalized. 

In the measured spectrogram from Fig.\ \ref{fig:4}(e), the fundamental modes of the depicted resonators are equalized in ERV/wavelength to within a standard deviation of about $0.5$ nm/$40$ pm, which can be considered as an upper bound on the detuning noise in this lattice. The post-processing of the device was focused on optimizing the depicted $q=3$ axial mode lattice. There is a substantial amount of uncertainty considering the adjacent couplings are of approximately the same order of magnitude or less. Despite this, the main topological feature of the lattice can easily be observed: the zero-energy topological mode (orange) that is present at the center and distributed is mostly on the sites adjacent to the domain wall (dashed line). Additionally, there is a distinct opening of auxiliary band gaps when crossing from the left to right sides. While the negative-energy interface bound mode (green) lies firmly below the bands on either side, the positive energy one is nearly at the same wavelength as the top of the band of the right, partly attributable to the Fano resonance lineshape and the lower uniformity of couplings for longer wavelength modes. Lastly, there exist resonances around the right lattice's auxiliary bandgaps (yellow) where the quasi-modes are expected to be, but they are close in energy with the bulk modes on the left and spatially merge into them, making these modes difficult to isolate.

\section{Theory \& Analysis of Multiband Heterojunctions} \label{sec:theo}

The topological nature of the interface between SSH models with dissimilar unit cells is only beginning to be explored \cite{zhao_interface_2024,mandal_topological_2024}, necessitating that we fill this gap in order to determine whether the modes observed in Fig.\ \ref{fig:4} are topological. The primary obstacle to generalizing standard topological theory to these systems is that the symmetries upon which topological invariants are defined are imperfect or only locally true, as opposed to being globally true. In other systems, imperfect symmetries have been dealt with by basing the definition of the topology on the scattering phases \cite{fulga_scattering_2012} of incident particles reflected from the system in question's boundaries. This definition can be employed in a significantly more general class of systems such as those with interparticle interactions or inexact symmetries as it only require determining the boundary Green's function \cite{peng_boundary_2017,gurarie_single-particle_2011,essin_bulk-boundary_2011}, which can be obtained through e.g., transfer-matrix \cite{wielian_transfer_2025,pineda_edge_2022} methods. However, current forms of this method are inherently focused on midgap zero-energy topological properties of external boundaries, which is only a part of the topology of the systems we are interested in. 

By basing the topology on relative phases of the Bloch modes, we can analyze the topological properties of systems that lack complete chiral and inversion symmetry (i.e., these symmetries are only present in the bulk) due to a non-integer number of unit cells or domain boundaries \cite{pletyukhov_surface_2020,pletyukhov_topological_2020,muller_universal_2021}.  In the bulk of a lattice (or for periodic boundary conditions), Bloch's theorem can be applied to group the eigenmodes of $\hat H$ into momentum modes $\ket k$. This folds the Hamiltonian into an $\ell \times \ell$ matrix $\hat H(k)=\bra{k}\hat H\ket{k}$ acting on the sublattice for each $k$ \cite{anastasiadis_bulk-edge_2022,lee_winding_2022}. Each of the $\ell$ eigenvectors of $\hat H(k)$ will define a band $\mu$ of eigenmodes $\ket{k_\mu}=\ket{u_\mu(k)}\otimes \ket{k}$, where we define these discrete Bloch modes and momentum eigenmodes as:
\begin{equation}
    \ket{u_\mu(k)}=\sqrt{\frac{1}{\ell}}\sum_{s=1}^\ell a_\mu(k,s)\exp\bigg[i\frac{k}{\ell}s-i \theta_\mu (k,s)\bigg]\ket{s}.
\end{equation}
with $\ket{k}=\sqrt\frac{2\pi \ell}{L}\sum_{n=n_1}^{n_L} \exp(i k n)\ket{n}$
The basis vectors are factored as $\ket{z_{n\ell+s-s_0}}=\ket{s}\otimes\ket{n}$ with sublattice index $s\in[1,\ell]$ and unit cell index $n\in[1,N]$ such that $s_j$ ($n_j$) is the sublattice (unit cell) index assigned to site $j$. Wherever $n\ell+s$ corresponds to a nonexistent site, the wavefunction is taken to be zero instead. The sublattice phases $\theta_\mu(k,s)$ and amplitudes $a_\mu(k,s)$ control the wavefunction's dependence on $s$.

Depending on the bulk's symmetries, isolated bound modes can appear within bandgaps and at the boundaries of SSH-like lattices. While SSH2 only allows modes within the central bandgap, non-zero energy modes can also emerge within the auxiliary bandgaps (those not centered at zero) when $\ell>2$, increasing the number of possible topological phases \cite{marques_one-dimensional_2019, 
anastasiadis_bulk-edge_2022}. When trying to form the eigenmodes of an open-boundary lattice by linearly combining Bloch modes \cite{anastasiadis_bulk-edge_2022}, the requirement that wavefunctions vanish just outside the lattice induces a condition that the phase of propagation between virtual sites $j=0$ and $j=L+1$ must be an integer multiple of $\pi$:
\end{multicols}
\begin{equation}\label{eqn:quantization}
    m_\mu(k)=\frac{\Big[kn_{L+1}+\frac{k}{\ell}s_{L+1}-\theta_\mu(k,s_{L+1})\\\Big]-\Big[kn_{0}+\frac{k}{\ell}s_0-\theta_\mu(k,s_0)\Big]}{\pi}\in \mathbb{Z}_N,
\end{equation}
\begin{multicols}{2}
Depending on the values of $s_{L+1}$ and $s_0$, the total number of unique solutions to Eq.\ \eqref{eqn:quantization} between all the bands may fall short of $L$, as the bound modes lie in the complex momentum plane at the time reversal invariant momenta $\mathrm{Re} \ k=\pm \pi$. We can therefore determine the number of bound modes from the band $Q$ by the difference between the total modes $L$ and the number of free modes (the modes within a reduced Brillouin zone \cite{anastasiadis_bulk-edge_2022}, equivalent to $1/2\int_{-\pi}^\pi m'_\mu(k) dk -1$ for each band), giving the formula:
\begin{equation}\label{eqn:Zak}
     Q=\ell-1+\sum_{\mu}\oint \frac{dk}{2\pi} \frac{\partial}{\partial k}\Big[\theta_\mu(k,s_{L+1})-\theta_\mu(k,s_0)\Big].
\end{equation}
This reduces to the winding number and Zak phase \cite{alexandradinata_wilson-loop_2014,guzman_geometry_2022,gurarie_single-particle_2011,asboth_short_2016} when $\ell=2$, and will be discretized in the general case because the discrete translational symmetry forces $\theta_\mu(+\pi,s)=\theta_\mu(-\pi,s)$ mod $2\pi$, regardless of other symmetries. Eq.\ \eqref{eqn:Zak} is gauge- and basis-independent, and can be considered a discrete version of Levinson's theorem \cite{ma_levinson_2006} composed of the previously defined ``normalized sublattice phases"  \cite{anastasiadis_bulk-edge_2022} with additional terms to remove the intracellular Zak phase \cite{rhim_bulk-boundary_2017}. Having the freedom of unit cell basis can reduce counting topological modes to counting the dangling sites on the edges of the lattice, as it is possible for all the normalized sublattice phases to be zero. All unit cell bases in this work are chosen this way except except in Appendix \ref{sec:SSH2/4}.

Converting Eq.\ \eqref{eqn:Zak} into the position basis, Gauss's law can be used to convert the expression from a sum over the boundaries to a sum over the bulk, making the total bound modes:
\end{multicols}\begin{equation}\label{eqn:Gauss}
    Q=\ell-1+\sum_{j=0}^{L} \Big[D(j+1)-D(j)\Big] \quad \mathrm{with} \quad D(j)=\sum_\mu\oint\frac{dk}{2\pi\ell} \frac{\braket{s_j|s_j+i\ell \partial_k|u_\mu(k)}}{\braket{s_j|u_\mu(k)}}.
\end{equation}\begin{multicols}{2}
The divergence of $D(j)$ averages to 0 across each unit cell, making changes in $D(j)$ across an integer number of unit cells signify the presence of bound modes. We will refer to $D(j)$ as the sublattice displacements, as it can be considered a generalization of chiral displacement/polarization \cite{guzman_geometry_2022,sridhar_measuring_2025} for non-chiral systems. It is a local \cite{cerjan_classifying_2024,cerjan_operator-based_2022} but gauge- and basis- independent property, making it serve as a convenient definition of the topological invariant for systems with local but not global symmetries such as the SSH2 - SSH4 junction of Fig.\ \ref{fig:4}. 

When the number of bands change between the left and right sides of a junction, the two sides do not naturally belong to the same topological classification system. Furthermore, we will see that non-topological bound modes -- modes that may disappear after mild deformations of the Hamiltonian -- appear, which motivates us to identify topological changes with the minimum number of bound modes and/or quasi-modes under topology-preserving variations (referred to as the net topological modes). We propose two methods of predicting the net topological modes in such scenarios. 

The first method is to guess and check topology-preserving \cite{asboth_short_2016} local deformations of the lattices to manually search for the minimal number of interface modes. Notably, unit cells can be formed or removed in groups of $\ell$ sites around the junction, potentially changing the number of interface modes by $\ell$ (e.g. Fig.\ \ref{fig:B5}(a) relative to Fig.\ \ref{fig:4}(a)), which is why the topological invariant can only be defined mod $\ell$. In a simple system like this, it is not difficult to guess a configuration that minimizes the number of modes- one can break up the pentamer at the junction into two dimers and a monomer. At this perturbed interface, the only mode is the zero-energy mode (Fig.\ \ref{fig:B4}(c)), implying that the 2 extra modes and quasi-modes apparent in Fig.\ \ref{fig:4}(b,c) are non-topological as the bulks have been left untouched.

We can more systemically approach the problem by bringing each part of the lattice under the same classification system and reconsidering both lattices as sharing a common period equal to their least common multiple (e.g.  $\ell=2$ and $\ell=4$ lattices are both $\ell=4$ lattices). The resulting systems lie directly at critical points, making eq.\ \eqref{eqn:Gauss} ill-defined. To deal with this we consider symmetry-preserving minute deformations of the Hamiltonian \cite{benalcazar_quantized_2017} to characterize the topology by the properties that persist on either side of the critical point (e.g., the sublattice displacements and bound modes).

Specifically, we take the first 4 cells of the lattice in Fig.\ \ref{fig:4}(a) as a unit cell of an SSH4 lattice with couplings $ v_L(j)=[1, 0.5, 1, 0.5]$, and then consider perturbations that split the auxiliary bands (Fig.\ \ref{fig:B5}) of the form ${v}_L'(j)=[1,0.5+\delta, 1, 0.5-\delta]$ (note that both the chiral and inversion symmetries of the bulk of the SSH4 model \cite{lee_winding_2022,mandal_topological_2024} are preserved). The original junction (when $\delta=0$) will have different interface modes in comparison to when $\delta>0$ or when $\delta<0$, as well as different sublattice displacements. Table \ref{tab:Bound mode table} compares the values of the sublattice displacement on a representative unit cell of the left lattice for $\delta<0$, $\delta=0$, and $\delta>0$ with the corresponding values on the right lattice and the minimum number topological modes as determined by the manual search method in Appendix \ref{sec:SSH2/4}. For the $\delta=0$ system, the changes in the sublattice displacements (computed in SSH2 basis) for each possible sublattice index do not give a consistent result: depending on how the lattice sites are assigned into groups, there can either be one or three interface modes, indicating a failure of the sublattice displacements to predict the number of topological modes. However, there are consistent predictions for the perturbed systems, albeit to two distinct common answers: the two possible predictions for the $\delta=0$ system. This seems to suggest that the inconsistency in the predictions for the $\delta=0$ system arises because it lies exactly at the critical point of SSH4 systems. Alternatively, it can be recognized that $D_L$ should only be considered to refer to unique phases mod 2, making the same hold for $D_L-D_R$, in which case the predictions of 1 and 3 modes become equivalent, making 1 the true number of net topological modes.
\end{multicols}
\begin{table}[h!]
\centering    
\begin{tabular}{cccccccc}\toprule
         $j_L$ $(j_R)$&   \multicolumn{3}{c}{$D_L$}&$D_R$&\multicolumn{3}{c}{$D_L-D_R$ mod 4} \\\midrule
 & $\delta<0$&$\delta=0$&$\delta>0$&&$\delta<0$& $\delta=0$&$\delta>0$ \\
 1 (17)& 3& 1&1&4&3& 1&1\\
 2 (18)& 4& 2&2&1&3& 1&1\\
 3 (19)& 1& 1&3&2&3& 3 (1)&1\\
 4 (20)&  2& 2&4&3&3& 3 (1)&1\\
 & & & \multicolumn{2}{c}{Topological Modes}& 3& 1&1\\ \end{tabular}
    \caption{Analysis of the number of topological modes using sublattice displacements of perturbed lattices. Sublattice displacements $D(j)$ on a representative unit cell ($j_L=1,2,3,4$) from the left lattice for $\delta=0$, $\delta<0$, and $\delta>0$ perturbations, as well as the corresponding ($j_R=17,18,19,20$) values on the right lattice. The last three columns indicate their differences mod 4 (mod 2) which gives the number of topological modes except at $\delta=0$. When $\delta=0$, only distinct values of $D_L$ mod 2 correspond to distinct topological modes, making $D_L-D_R$ only correspond to unique phases mod 2, which is why the number of interface modes is written as 3 (1).}
    \label{tab:Bound mode table}
\end{table}
\vspace{-40pt}
\begin{multicols}{2}
Both methods -- manually testing local deformations and computing sublattice displacements -- have thus concluded that there must be a bound mode at the interface within the central bandgap, corresponding to a change in the topology across said interface. While the former takes a heuristic path contingent on guessing the result of changing certain couplings, the latter can be employed as a brute-force solution to determining the possible configurations of the interface and the net number of topological modes. Alternatively, the sublattice displacements can educate and affirm guesses of local deformations minimizing the number of interface modes, making the combination of the two methods a versatile and reliable approach to analyzing the topological properties of lattice heterojunctions that posses multiple bands. While this method was only applied to a 1D system, it can likely be extended to higher-dimensional systems akin to how chiral methods \cite{gurarie_single-particle_2011} extend from 1D to higher dimensions. 
\section{Conclusion}\label{sec12}
In summary, we have identified how the ultra-low loss and high fabrication precision of the SNAP platform make it substantially promising for 1D analogue Hamiltonian simulation. We demonstrate its unique capability to achieve non-volatile matching between more than 20 non-uniformly coupled resonator sites. The platform naturally supports multiple axial mode orders at each site, meaning our devices host an intrinsic parameter sweep over 5 distinctly-behaving coupling configurations both far and close to the central topological critical point. We not only demonstrate the first topological lattices on the SNAP platform, but do so with angstrom-level precision. Following this, we ventured beyond what has been accomplished on other platforms
by experimentally investigating multiband topological interfaces. To analyze this more general class of topological systems, we developed a methodology based on extending notions of topological charge and polarization to lattice models with arbitrary numbers of bands and without global symmetries. Notably, this analysis revealed that the structure possessed both a topological interface mode as well as two non-topological interface modes per axial mode lattice. 

Our work lays a foundation for exploration of topological light control and its robustness to system disorder \cite{Cerjan:20}, with exciting future extensions of this work involving cavity-enhanced nonlinear-optical effects to couple resonant frequency modes \cite{Eadie:25,Crespo:23,Kolesnikova:23}. This would unlock additional degrees of freedom (polarization, orbital angular momentum, etc.) as synthetic dimensions \cite{yu_comprehensive_2025} with which 2 or higher-dimensional topology can be investigated. Likewise, the theoretical topological framework developed in this work could potentially be extended to analyze such higher-dimensional systems as well as to analyzing topological insulators lacking symmetries in general.
\vspace{-3pt}\bmhead{Acknowledgments} The authors would like to thank Dr. Fredrik Fatemi for support in producing the tapered microfibers used in this work. The authors also thank Chad Smith for creating Fig.\ \ref{fig:1}(a). A.D. acknowledges support from the National Science Foundation (NSF) CAREER award (\# 2340835). N.F. acknowledges support from NSF (QuSeC-TAQS \# 2326792) and Army Research Laboratory through the NSF-DEVCOM INTERN Program and from the DoD SMART Scholarship Program.
\end{multicols}

\bigskip

\begin{appendices}
\begin{multicols}{2}
\section{Fiber Tapering}\label{sec:taper}
A Vytran GPX-3800 Glass Processing System system was used to taper SMF-28 fibers to a diameter of approximately 1 $\mu$m to ensure only a single mode could be supported. The tapering processes was broken into two steps: initially a 2 mm section of the fiber was heated and then stretched apart by 9 mm on each side, followed by heating a 4 mm region of the fiber and stretching by 8 mm on each side, resulting in an increase in the fiber length by 34 mm. 
\end{multicols}
\begin{figure}[h]
    \centering
    \includegraphics[width=1\linewidth]{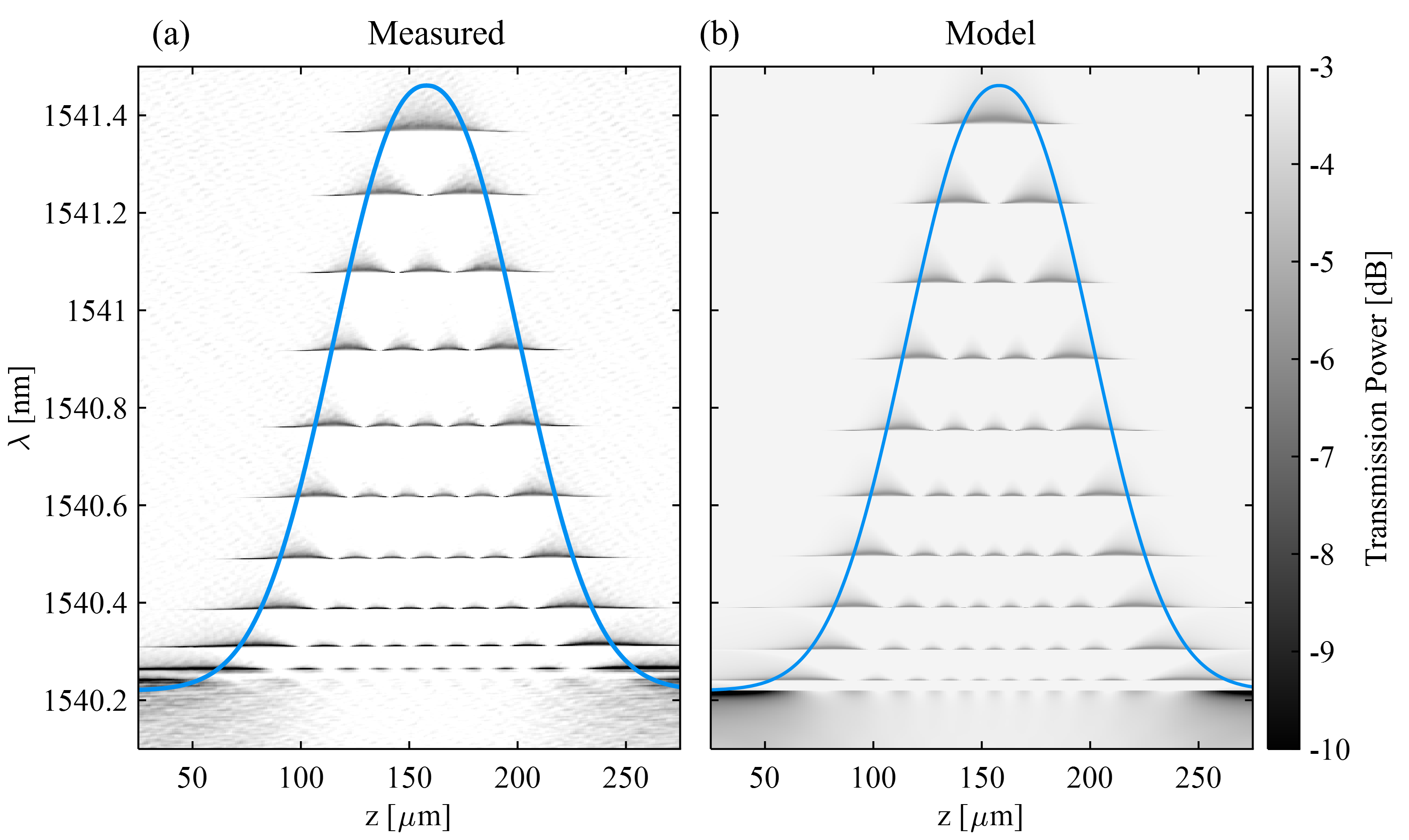}
    \caption{\textbf{(a)} Measured and \textbf{(b)} Fitted Model Spectrogram of a SNAP resonator comparable to those created in this work.}
    \label{fig:B1}
\end{figure}
\begin{multicols}{2}
\section{SNAP Spectrum Fitting \& Modeling}\label{sec:fit}
\figref{fig:B1} shows a measured and model spectrogram of a lone SNAP resonator. The model is obtained by assuming a super-Gaussian ERV curve:
 \begin{equation}
     \Delta r(z)=\frac{\bar r}{\bar \lambda}\Bigg[\Lambda \exp\bigg(-\Big(\frac{z-z_0}{\sqrt{2}\sigma}\Big)^{2p}\bigg)+h\Bigg],
 \end{equation}
where $\Lambda$ represents the height  of the resonator along the  wavelength axis, $z_0$ its central location, $\sigma$ its width, and $p$ is a factor representing the ``sharpness" of the ERV curve in comparison to a normal gaussian curve. $\bar r$ and $\bar \lambda$ are defined as before, and $h$ accounts for errors in the estimate of $\bar \lambda$. These parameters are fit by finding the bare resonant wavelengths $\Delta \lambda_q^{\mathrm{meas}}$ by locating the transmission minimum within the spectrogram (specifically at their caustics to avoid the effects of the taper renormalization \cite{Vitullo:20}), then determining the eigenvalues of Eq.\ \eqref{eqn:1Dschrodinger} from this potential corresponding to the bound modes $\Delta\lambda_q^{\mathrm{model}}$. Then a gradient-descent minimization of the listed parameters with a cost function equal to sum of the squared differences of $\Delta \lambda_q^{\mathrm{meas}}$ and $\Delta \lambda_q^{\mathrm{model}}$ for $q\in [1,10]$ can be employed. In Fig.\ \ref{fig:B1}(b), the result is that $\Lambda=1.240$ nm, $\sigma=39.7$ $\mu$m, and $p=1.117$.

\begin{figure}
    \centering
    \includegraphics[width=0.8\linewidth]{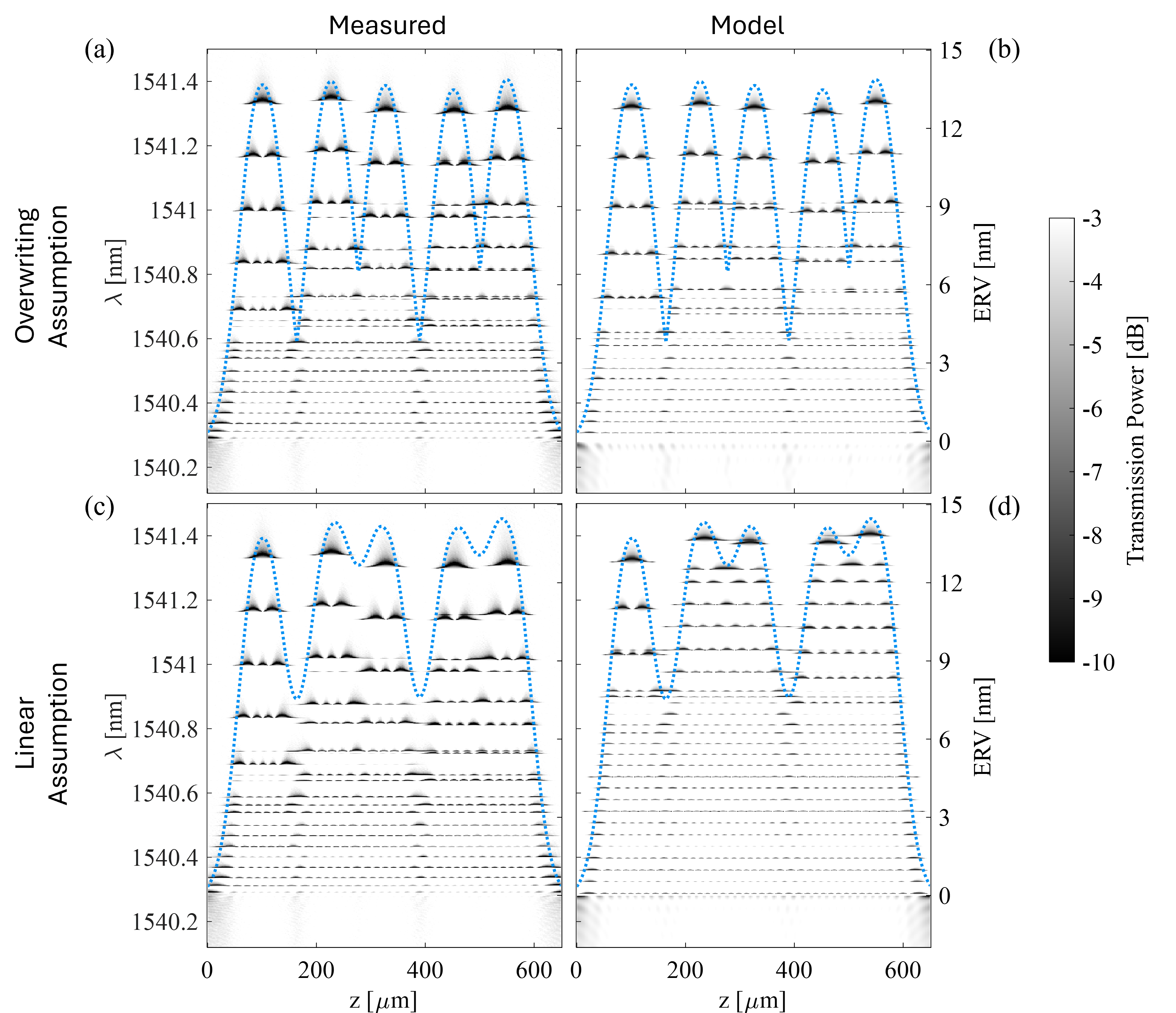}
    \caption{Comparison of modeling approaches for 5-site SSH model shown in Fig. 3. Measured spectrograms of SNAP device from Fig.\ \ref{fig:3} (a,c) where each resonator has an ERV curve based on fit parameters of Fig.\ \ref{fig:B1} but with $\Lambda=1.11$ nm and different assumptions about how ERV from separate exposures combine. In (a) it is assumed that the ERV of the device abides by eq.\ \eqref{eqn:max} whereas (c) assumes ERV growth abides by eq.\ \eqref{eqn:lin}. The corresponding computed spectrograms with the same ERV curves are depicted in (b) and (d) respectively. The ERV curve in (c) fails to follow the edges of the mode profiles, and spectrogram (d) does not resemble the measured ones in (a) and (c) as a whole, indicating a failure of the linear assumption of ERV growth. }
    \label{fig:B2}
\end{figure}

When multiple SNAP resonators labeled by position index $j$, each individually corresponding to a super-Gaussian ERV profile $\Delta r_j(z)$, are fabricated adjacent to each other, the modes of the overall structure are determined by some cumulative ERV profile $\Delta r(z)$. Due to the effects of the nonlinearity of the CO$_2$ annealing process on positions that receive multiple contributions, it will not generally be the case that multiple identical laser pulse exposures at nearby axial positions will give rise to a net ERV profile equal to the sum of individual super-Gaussian ERV curves. This means we find the cumulative ERV at any location more closely resembles the \textit{maximum} of the individual ERV profiles at any given location. The ERV profiles are better modeled as ``overwriting" each other,
\begin{equation}\label{eqn:max}
    \Delta r(z)\approx \max_j\Delta r_j(z),
\end{equation}
as opposed to 
\begin{equation}\label{eqn:lin}
    \Delta r(z)=\sum_j \Delta r_j(z).
\end{equation}
Fig.\ \ref{fig:B2} compares ERV profile models given by Eq.\ \eqref{eqn:max} and Eq.\ \eqref{eqn:lin} by applying them to the measured 5-site SSH2 lattice (analyzed in Fig.\ \ref{fig:3}). In each diagram, the $\sigma$ and $p$ parameters are taken from the results of fitting the lone resonator in Fig.\ \ref{fig:B1}, but $\Lambda$ has been rescaled to better represent the size of these resonators (i.e.\ they are reduced from 1.24 nm to the values shown in Table\ \ref{tab:Fit}). The ERV curve shown in Figs.\ \ref{fig:B2}(a) and \ref{fig:B2}(b) are computed under the overwriting model (Eq.\ \eqref{eqn:max}), whereas the ERV curve of (c) and (d) are computed under the linear model (Eq.\ \eqref{eqn:lin}). The ERV profile using the overwriting profile gives rise to a modal spectrum that is a much better match to the measured data than the linear model. 

In this situation, is is clear the overwriting assumption is both the superior model and an accurate one, motivating its use in fitting for Fig. \ref{fig:3}. The ERV profile for each resonator was permitted to have a distinct width $\sigma$ and $\Lambda$, but a common $p$ value. The results are $p=1.144$ and the values for $\Lambda$ and $\sigma$ are written in Table \ref{tab:Fit}.

The similarity between these fitted widths and the $p$ value to that of the individual resonator in Fig.\ \ref{fig:B1} further affirms the validity the of the ERV-overwriting assumption. 

It should be noted that the overwriting model underestimates ERV growth in situations where the intersite spacing $\Delta z$ is comparable to the mode widths $\Delta w_q$ (and/or the resonator widths $\sigma$). This can be easily verified by recognizing that the overwriting model implies there would be no ERV change at all during the post-processing steps in which the resonances are matched. 
\end{multicols}\begin{table}
    \centering
     \begin{tabular}{c|ccccc}\toprule
         Resonator \# &  1&  2&  3&  4& 5\\\midrule
         $\Lambda$ (nm)&  1.108&  1.119&  1.106&  1.093& 1.124\\
         $\sigma$ ($\mu$m)&  37.2&  39.6&  35.8&  37.0& 38.7\\ \bottomrule
    \end{tabular}
    \caption{Fitted values of parameters for each of the resonators in Fig.\ \ref{fig:3}. Every resonator had a common value of $p$ fitted, for which the result was $p=1.144$.}
    \label{tab:Fit}
\end{table}
\begin{multicols}{2}
\section{SSH2/SSH4 Heterojunction Comparisons} \label{sec:SSH2/4}
\begin{figure}[H]
    \centering
    \includegraphics[width=1\linewidth]{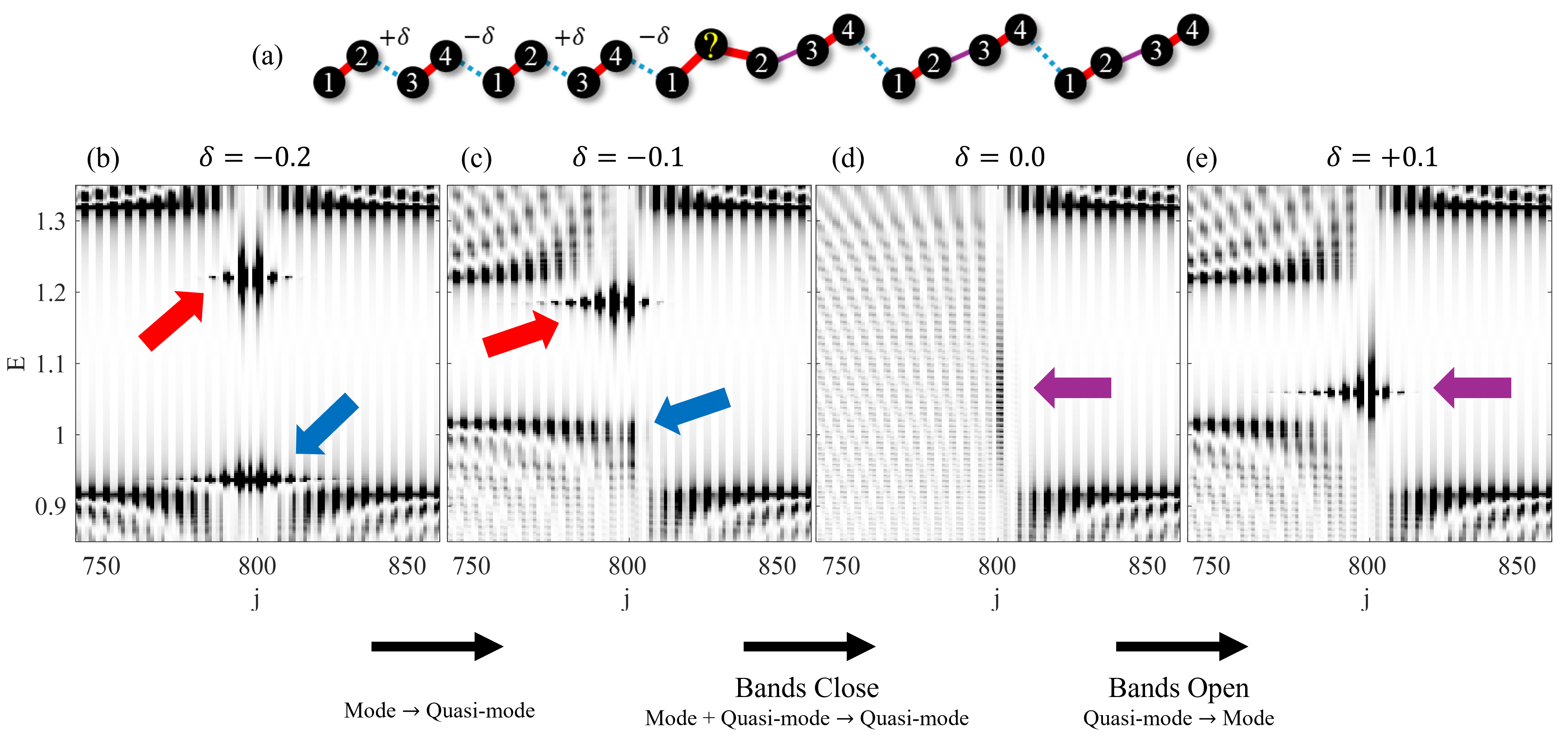}
    \caption{\textbf{(a)} Polymer model of the perturbed lattice heterojunction, and (b-e) partial views of the simulated bare tight-binding Green's functions (arb. units for intensity) showing how the bare mode structure varies with $\delta$ across the critical point at $\delta=0$ . \textbf{(b)} $\delta=-0.2$ Two real modes (red and blue arrows) are visible within the gap. \textbf{(c)} $\delta=-0.1$ The lower of the two modes (blue arrow) from is converted into a quasi-mode as it enters the band. \textbf{(d)} $\delta=0$ The closing of the bandgap cause the quasi-mode and real mode to merge into a single quasi-mode (purple arrow). \textbf{(e)} $\delta=+0.1$ The quasi-mode becomes a real mode (purple arrow) after the bandgap opens again.}
    \label{fig:B4}
\end{figure}

Fig.\ \ref{fig:B4} contains close-up views of the auxiliary bandgaps of heterojunctions like that of Fig.\ \ref{fig:4}, but with the $\delta$ perturbations as described in Sec.\ \ref{sec:theo}, such that the couplings are $v_L=[1, 0.5+\delta, 1, 0.5-\delta]$ with $N_L=200$ on the left (formerly SSH2) side, and $v_R=[1, 0.7, 1, 0.3]$ with $N_R=200$ on the right side (Fig.\ \ref{fig:B4}(a)). A nonzero $\delta$ opens a bandgap in the middle of each of the SSH2 bands, placing it on similar footing with SSH4. For all values of $\delta$ there is a zero-energy mode and the pair of modes above and below all the bands like in Fig.\ \ref{fig:4}(b-c), which is why only the auxiliary bandgap is shown. For $\delta=-0.2$  (when the bandgaps of the left side and right side are equal, and the system has exact inversion symmetry about the interface), (Fig.\ \ref{fig:B4}(b)) there are two bound modes (red and blue arrows) present in the auxiliary band gap. However, (Fig.\ \ref{fig:B4}(c)) for $\delta<0$  except $\delta\neq-0.2$, the lower (blue) of the two modes becomes a quasi-mode, as its energy falls within the bandgap (For $\delta<-0.2$, the spectrograms are formally equivalent to Fig.\ \ref{fig:B4}(c) but with the left and right sides switched). When $\delta$ reaches zero as the auxiliary bandgaps close (Fig.\ \ref{fig:B4}(d)), the quasi-mode and real mode combine into a single quasi-mode (purple arrow). Past (Fig.\ \ref{fig:B4}(e)) $\delta>0$, there is only a single bound mode within the auxiliary bandgap, implying that $\delta=0$ is a critical point of the system. 
\end{multicols}\begin{figure}[H]
    \centering
    \includegraphics[width=1\linewidth]{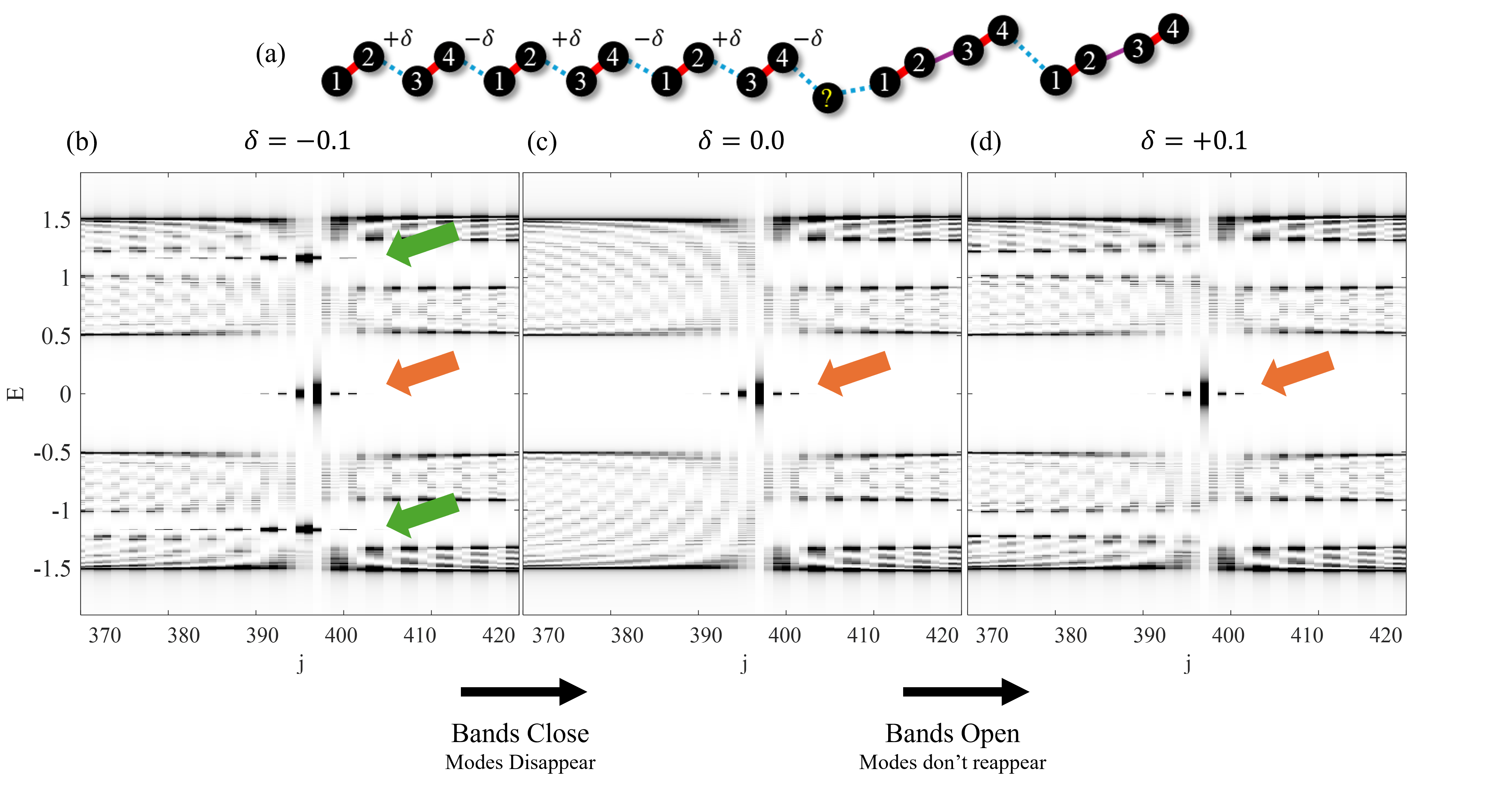}
    \caption{\textbf{(a)} Locally deformed and perturbed lattice. (b-d) bare tight-binding Green's functions of systems that have been perturbed by $\delta$, but have also been deformed to possess the minimum number of interface modes. \textbf{(b)} For $\delta<0$, there are three modes- a zero energy bound mode (orange), and two modes within the auxiliary bandgaps (green). \textbf{(c)} For $\delta=0$, the auxiliary bound modes disappear, and don't reappear for (d) $\delta>0$.}
    \label{fig:B5}
\end{figure}\begin{multicols}{2}
We also examine the $\delta$-perturbed heterojunction after topology-preserving local deformations (e.g. Fig.\ \ref{fig:B5}(a)) to find the minimum number of interface modes -- a topologically robust quantity by definition with which we can compare the number of apparent modes (Fig. \ref{fig:B4}) and the sublattice displacements. Notably, the zero-energy mode continues to appear in every case (Fig.\ \ref{fig:B5}(b-d)), and quasi-modes fail to appear in any case. At the $\delta=0$ (Fig.\ \ref{fig:B5}(c)) band closing point, the interface transitions from hosting 3 bound modes ($\delta<0$, Fig.\ \ref{fig:B5}(b)) to having only a single bound mode (Fig.\ \ref{fig:B5}(d)), re-affirming $\delta=0$ as a critical point.

We list the real and quasi-modes for each case depicted in Fig.\ \ref{fig:B4} in Table \ref{tab:Bound modes appendix}, along with their sum mod 4 (the ``total" mode count) and the sublattice displacements from Table\ \ref{tab:Bound mode table}. The total net mode count is equal to 3 for $\delta<0$ and 1 otherwise, suggesting that $\delta=0$ is a critical point. For all values of $\delta$, the minimum mode count matches the total mode rather than the real mode count (mod 4), suggesting that quasi-modes are a valid indicator of topological phase changes just like real modes (i.e., quasi-modes are a form of topological ``charge"). Likewise, the sublattice displacements \textit{almost} always line up with the net and minimum mode count, with the critical point $\delta=0$ as the only exception. The correct result can be obtained by recognizing that the predicted number of modes should only be unique mod 2 for $\delta=0$ (since the left side has two bands), or by taking the minimum of all the predictions across the sublattice. 
\end{multicols}\begin{table}[h]
    \centering \begin{tabular}{c|cccc}\toprule
         Mode Count&$\delta=-0.2$&  $\delta\neq-0.2$ \& $\delta<0$&  $\delta=0$& $0<\delta$\\\midrule
         Real&7&  5&  3& 5\\
         Quasi-&0&  2&  2& 0\\
 Total& 3& 3& 1&1\\
$D_L-D_R$&3& 3& 3/1&1\\ 
 Minimum& 3& 3& 1&1\\ \bottomrule
    \end{tabular}
    \caption{Tabulation of modes and quasi-modes observed in spectrograms from Fig.\ \ref{fig:B4}(b-d), with the total modes (sum of the modes and quasi-modes mod 4). Minimum modes consider the lowest number of bound modes localized to the interface under local deformations like in Fig.\ \ref{fig:B5}. Additionally, the minimum of the changes in the sublattice displacement across the sublattice from Table\ \ref{tab:Bound mode table} are provided. In all cases, the net modes coincide with the minimum modes and the minimum of the displacement across the sublattice. The $\delta=0$ point is a critical point between the system having 1 or 3 net topological modes.}
    \label{tab:Bound modes appendix}
\end{table}
\backmatter
\end{appendices}

\bibliography{avikrefs,NateZot,snap_references,sn-bibliography}

\end{document}